\documentstyle[aps,pre,manuscript]{revtex}

\begin{document}

\draft

\title{
An Analytical and numerical investigation of escape rate for a noise
driven bath
}

\author{Jyotipratim~Ray~Chaudhuri\footnote{present address : Department of
Physics, Katoa College, Katoa, Burdwan, West Bengal, India.},
Suman~Kumar~Banik, Bidhan~Chandra~Bag
and Deb~Shankar~Ray
\footnote{\tt e-mail : pcdsr@mahendra.iacs.res.in} }

\address{Indian Association for the Cultivation of Science, Jadavpur,
Calcutta 700 032, India.}

\date{\today}

\maketitle

\begin{abstract}
We consider a system-reservoir model where the reservoir is modulated by an
external noise. Both the internal noise of the reservoir and the external
noise are stationary, Gaussian and are characterized by arbitrary decaying
correlation functions. Based on a relation between the dissipation of the
system and the response function of the reservoir driven by external noise 
we numerically examine the model using a full bistable potential to show that
one can recover the turn-over features of the usual Kramers' dynamics when
the external noise modulates the reservoir rather than the system directly.
We derive the generalized Kramers' rate for this nonequilibrium open system.
The theoretical results are verified by numerical simulation.
\end{abstract}

\vspace{0.5cm}

\pacs{PACS number(s) : 05.40.-a, 02.50.Ey}

\section{ Introduction 
\label{sd1} 
}

More than half a century ago Kramers proposed a diffusion model for
chemical reactions in terms of the theory of Brownian motion in phase space
\cite{Kramers}.
Since then the model and several of its variants have been ubiquitous in
many areas of physics, chemistry and biology for understanding the nature
of activated processes in 
classical \cite{Van,Grote-Hynes,Hanggi,Pollak,JRC-1,Banik},
quantum and semiclassical \cite{Caldeira,Grabert-1,Grabert-2,JRC-2}
systems, in general. These
have become the subject of several reviews \cite{RMP,Melnikov,Talkner} 
and monograph \cite{Weiss} in the recent past.

In the majority of these treatments one is essentially concerned with an
equilibrium thermal bath at a finite temperature which stimulates the 
reaction co-ordinate to cross the activation energy barrier. The inherent
noise of the medium is of internal origin. This implies that the dissipative 
force which the system experiences in course of its motion in the medium
and the stochastic force acting on the system as a result of random impact
from the constituents of the medium arise from a common mechanism. From a 
microscopic point of view the system-reservoir Hamiltonian description
\cite{Ford,Louisell,Zwanzig,Lindenberg}
developed over the decades suggests that the coupling of the system and
the reservoir co-ordinates determines both the noise and the dissipative 
terms in the Langevin equation describing the motion of the system. It is 
therefore not difficult to anticipate that these two entities get related
through a fluctuation-dissipation relation \cite{Kubo} 
( these systems are sometimes classified as thermodynamically closed system 
in contrast to the systems driven by external noise in 
nonequilibrium statistical mechanics \cite{West} ). However,
when the reservoir is modulated by an external noise it is likely that this
relation gets affected in a significant way.
Since  the modulation of the reservoir crucially
depend on its response function, one can further envisage a connection 
between the dissipation of the system and the response function of the 
reservoir due to the external noise from a microscopic point of view.

In the present paper we explore this connection in the context of activated
rate processes when the reservoir is modulated by an external noise. 
Specifically our object here is twofold: 
(i) to explore the role of reservoir response as a function of external noise
on the system dynamics in contrast to direct driving of the system by external 
noise,
(ii) to calculate the generalized Kramers' rate for the steady state of this 
nonequilibrium open system taking full care of thermodynamic consistency 
condition. 
Both the internal and the external noises are
Gaussian, stationary and are characterized by arbitrary decaying correlation
functions. While the internal noise of the reservoir is thermal, the external 
noise may be of
thermal or non-thermal type. We consider the stochastic motion to be spatial
diffusion limited and calculate the rate of escape in the intermediate to
strong damping regime. It is worth to mention that the externally generated
nonequilibrium fluctuations can bias the Brownian motion of a particle in an
anisotropic medium and may be used for designing molecular motors and pumps
\cite{Astumian}. We further mention that nonequilibrium,
non-thermal systems have also been investigated phenomenologically by
a number of workers in several other contexts, e.g., for examining the role
of color noise in stationary probabilities \cite{Rattray}, 
properties of nonlinear systems \cite{Moss}, nature of crossover \cite{Jaume}, 
rate of diffusion limited coagulation processes \cite{Werner}, 
effect of monochromatic noise \cite{SJBE}, etc. While these treatments
concern direct driving of the system by an external noise the present
consideration is based on modulation of the bath.
A number of different situations depicting the modulation of the bath
by an external noise may be physically relevant. As, for example, we consider
a simple unimolecular conversion (say, an isomerization reaction) from
$A \longrightarrow B$. The reaction is carried out in a photochemically 
active solvent under the influence of external fluctuating light intensity.
Since the fluctuations in the light intensity result in the fluctuations in
the polarization of the solvent molecules, the effective reaction field
around the reactant system gets modified. Provided the required stationarity 
of this nonequilibrium open system is maintained (which is not difficult 
in view of the experiments performed in the studies of external noise-induced 
transitions in photochemical systems \cite{WH-RL}) the dynamics of barrier
crossing becomes amenable to the present theoretical analysis that follows.

The remaining part of this paper is organized as follows: In Sec.~{II} we
discuss a system-reservoir model where the later is modulated by an external
noise and establish an interesting connection between the dissipation of
the system and the response function of the reservoir due to external noise.
The stochastic motion in a linearized potential field is described in terms
of a Fokker-Planck equation in Sec.~{III}. Based on the traditional flux
over population method \cite{Farkas} we derive in Sec.~{IV} the generalized
expression for the Kramers' rate of escape from a metastable well. In 
Sec.~{V} we numerically analyze the model and the bath modulated dynamics
for the full potential and verify the theoretical rate with numerical
simulation. The paper is concluded in Sec.~{VI}.

\section{ The system-reservoir model : The reservoir modulated by  an 
external noise 
\label{sd2}
}

We consider a classical particle of mass $M$ is 
linearly coupled to a heat bath of $N$ harmonic oscillators driven by an
external noise. The total Hamiltonian is given by \cite{Zwanzig}
\begin{equation}
\label{eq1}
H = \frac{p^2}{2M} + V(x) + \frac{1}{2} \sum_{i=1}^N \left \{ 
\frac{p_i^2}{m_i} + m_i \omega_i^2 (q_i - g_i x )^2 \right \} + H_{int}
\; \; .
\end{equation}

\noindent
In Eq.(\ref{eq1}), $x$ and $p$ are the co-ordinate and momentum of the system
particle; $(q_i, p_i)$ are the variables associated with the $i$-th oscillator
and $\omega_i$ and $m_i$ are the corresponding frequency and mass,
respectively. $g_i x$ measures the
interaction between the particle and the bath. $V(x)$ is the potential
energy of the particle. $H_{int}$ is assumed to be of the form
\begin{equation}
\label{eq2}
H_{int} = \frac{1}{2} \sum_{i=1}^N \kappa_i \; q_i \; \epsilon (t) \;\; .
\end{equation}

\noindent
The coupling function $\kappa_i$ measures the strength of interaction and 
$\epsilon (t)$ is the external noise which we assume to be stationary and
Gaussian with zero mean, i.e., $\langle \epsilon (t) \rangle_e = 0$ and is
characterized by an arbitrary correlation function as follows;
\begin{eqnarray*}
\langle \epsilon (t) \epsilon (t') \rangle_e = 2 D \Psi (t-t') \; \; .
\end{eqnarray*}

\noindent
Here $\langle \ldots \rangle_e$ implies the averaging over the external 
noise.

We then eliminate the bath degrees of freedom in the usual way 
\cite{Ford,Zwanzig,Lindenberg,Bravo} to obtain 
the following generalized Langevin equation
\begin{eqnarray}
\label{eq4}
\dot{x} & = & v \; \; ,\nonumber \\
\dot{v} & = & -\frac{dV}{dx} - \int_0^t dt' \; \gamma (t-t') \; v(t') + f (t)
+ \pi (t) 
\end{eqnarray}

\noindent
[ while constructing Eq.(\ref{eq4}) we have set $M$ and $m_i$ equal to
unity ] where
\begin{equation}
\label{eq5}
\gamma (t) = \sum_{i=1}^N g_i^2 \; \omega_i^2 \; \cos \omega_i t \; \; .
\end{equation}

\noindent
$f(t)$ is the internal forcing function generated through the coupling 
between the system and the heat bath and is given by
\begin{equation}
\label{eq6}
f (t) = \sum_{i=1}^N g_i \; \left \{ \left [ q_i (0) - g_i x(0) \right ]
\; \omega_i^2 \; \cos \omega_i t + v_i (0) \; \omega_i \; \sin \omega_i t
\right \} \; \; .
\end{equation}

\noindent
In Eq.(\ref{eq4}), $\pi (t)$ is a fluctuating force term due to the external 
noise $\epsilon (t)$ and is given by
\begin{equation}
\label{eq8}
\pi (t) = - \int_0^t \varphi (t-t') \; \epsilon (t') \; dt' \; \; ,
\end{equation}

\noindent
where
\begin{equation}
\label{eq9}
\varphi (t) = \sum_{i=1}^N g_i \; \kappa_i \; \omega_i \sin \omega_i t \; \; .
\end{equation}

The form of Eq.(\ref{eq4}) therefore suggests that the system is driven by 
two forcing functions $f(t)$ and $\pi (t)$. $f(t)$ depends on the initial
conditions of the bath oscillators for a fixed choice of the initial 
condition of the system degrees of freedom. To define the statistical
properties of $f(t)$, we assume that the {\it initial distribution} is one in
which the bath is equilibrated at $t=0$ in the {\it presence} of the system 
but in the {\it absence} of the external noise agency such that
$\langle f(t) \rangle = 0$ and 
$\langle f(t) f (t') \rangle = k_BT \gamma (t-t')$.

Now, at $t=0_+$, the external noise agency is switched on and the bath
is modulated by $\epsilon (t)$. The system is governed by Eq.(\ref{eq4}),
where, apart from the internal noise $f(t)$, another fluctuating force
$\pi (t)$ appears, that depends on the external noise $\epsilon (t)$.
Therefore, one can define an effective noise $\xi (t) ( = f (t) + \pi (t) )$ 
whose correlation is given by
\begin{equation}
\label{neq6}
\langle \langle \xi (t) \xi (t') \rangle \rangle = k_BT \gamma(t-t')
+ 2D \int_0^t dt' \int_0^{t'} dt'' \; \varphi (t-t') \varphi (t'-t'')
\Psi (t'-t'') \; \; ,
\end{equation}

\noindent
along with $\langle \langle \xi (t) \rangle \rangle = 0$, where 
$\langle \langle \ldots \rangle \rangle$ means we have taken two averages
independently.  
It should be emphasized that the above relation (\ref{neq6}) is not a 
fluctuation-dissipation relation due to the appearance of the external noise
intensity. Rather it serves as a thermodynamic consistency condition.

Let us now digress a little bit about $\pi (t)$.
The statistical properties of $\pi (t)$ are determined by the normal mode 
density of the bath frequencies, the coupling of the system with the bath, 
the coupling of the bath with the external noise and the external noise 
itself. Eq.(\ref{eq8}) is reminiscent of the familiar linear
relation between the polarization and the external field where $\pi$ and
$\epsilon$ play the role of the former and the later, respectively. 
$\varphi (t)$
can then be interpreted as a response function of the reservoir due to external
noise $\epsilon (t)$. The very structure of $\pi (t)$ suggests that this
forcing function although originating from an external force is different
from a direct driving force acting on the system. The distinction lies at
the very nature of the bath characteristics (rather than system
characteristics) as reflected in the relation (\ref{eq8}) and (\ref{eq9}).

With the coupling coefficients
$g(\omega ) = g_0/\sqrt{\tau_c} \omega$ and 
$\kappa (\omega ) = \sqrt{\tau_c} \omega  \kappa_0$,
in the continuum limit \cite{Bravo}
$\gamma (t)$ and $\varphi (t)$ reduce to the 
following forms
\begin{equation}
\label{eq12}
\gamma (t)  =  \frac{g_0^2}{\tau_c} \;
\int d\omega \; {\cal D} (\omega) \; \cos \omega t 
\end{equation}

\noindent
and
\begin{equation}
\label{eq13}
\varphi (t)  =  g_0 \; \kappa_0 \; \int d\omega \; {\cal D} (\omega) \; 
\omega \; \sin \omega t \; \; .
\end{equation}

\noindent
where $g_0$ and $\kappa_0$ are constants and $\tau_c^{-1}$ is the 
cutoff frequency of the oscillator. ${\cal D} (\omega)$ is the density
of modes of the heat bath.

\noindent
From the above two relations, we obtain
\begin{equation}
\label{eq14}
\frac{d\gamma}{dt} = -\frac{g_0}{\kappa_0} \; \frac{1}{\tau_c} \varphi (t) \; \; .
\end{equation}

\noindent
Eq.(\ref{eq14}) is an important content of the present model. This expresses
how the dissipative kernel $\gamma (t)$ depends on the response function
$\varphi (t)$ of the medium due to external noise $\epsilon (t)$
[ see Eq.(\ref{eq8}) ]. Such a relation for the open system can be anticipated
in view of the fact that both the dissipation and the response function 
crucially depend on the properties of the reservoir especially on its density
of modes and its coupling to the system and the external noise source.
In what follows we shall be concerned with the 
consequences of this relation in terms of the Langevin description in the 
next section ( Eq.(\ref{eq15}) ) and numerical analysis of the full model
potential in Sec.~{V}.

\section{ The Fokker-Planck description of the linearized motion : 
Asymptotic analysis of the Fokker-Planck coefficients
\label{sd3}
}

We now consider the system to be a harmonically bound particle of unit mass
and of frequency $\omega_0$. Then because of Eq.(\ref{eq8}) the Langevin 
equation (\ref{eq4}) becomes
\begin{eqnarray}
\label{eq15}
\dot{x} & = & v \; \; , \nonumber \\
\dot{v} & = & -\omega_0^2 x - \int_0^t dt' \; \gamma (t-t') \; v(t') + f (t)
- \int_0^t dt' \; \varphi (t-t') \; \epsilon (t')
\end{eqnarray}

\noindent
The Laplace transform of Eq. (\ref{eq15}) allows us to write a formal 
solution for the displacement of the form
\begin{eqnarray}
\label{eq16}
x (t) & = & \langle \langle x (t) \rangle \rangle + 
\int_0^t dt' \; h (t-t')\; f (t') -
\frac{\kappa_0}{g_0}\tau_c \omega_0^2 \int_0^t dt' \; h (t-t')\; \epsilon (t')
\nonumber \\
& & - \frac{\kappa_0}{g_0}\tau_c \int_0^t dt' \; h_2 (t-t')\; \epsilon (t')
\; \; ,
\end{eqnarray}

\noindent
where we have made use of the relation (\ref{eq14}) explicitly.

\noindent
Here
\begin{equation}
\label{eq17}
\langle \langle x (t) \rangle \rangle = \chi_x (t) x (0) + h (t) v (0)
\end{equation}

\noindent
with $x(0)$ and $v(0)$ being the initial position and initial velocity of the 
oscillator, respectively, which are nonrandom and
\begin{equation}
\label{eq18}
\chi_x (t) = \left [ 1 - \omega_0^2 \int_0^t h (\tau) \; d\tau \right ] 
\; \; .
\end{equation}

\noindent
The kernel $h(t)$ is the Laplace inversion of 
\begin{equation}
\label{eq19}
\tilde{h} (s) = \frac{1}{s^2 + \tilde{\gamma} (s) \; s + \omega_0^2}
\end{equation}

\noindent
where, $\tilde{\gamma} (s) = \int_0^\infty e^{-st} \; \gamma (t) \; dt$, is 
the Laplace transform of the friction kernel $\gamma (t)$, and
\begin{equation}
\label{eq20}
h_2 (t) = \frac{d^2 h(t)}{dt^2} \; \;.
\end{equation}

\noindent
The time derivative of Eq.(\ref{eq16}) yields
\begin{eqnarray}
\label{eq21}
v (t) & = & \langle\langle v (t) \rangle \rangle 
+ \int_0^t dt' \; h_1 (t-t')\; f (t') -
\frac{\kappa_0}{g_0}\tau_c \omega_0^2 \int_0^t dt' \; h_1 (t-t')\; \epsilon (t')
\nonumber \\
& & - \frac{\kappa_0}{g_0}\tau_c \int_0^t dt' \; h_3 (t-t')\; \epsilon (t')
\end{eqnarray}

\noindent
where
\begin{equation}
\label{eq22}
\langle\langle v (t) \rangle \rangle = -\omega_0^2 h(t) + v (0) h_1 (t) \; \; ,
\end{equation}

\begin{equation}
\label{eq23}
h_1 (t) = \frac{d h(t)}{dt}  \; \; {\rm and} \; \;
h_3 (t) = \frac{d^3 h(t)}{dt^3} \; \;.
\end{equation}

Next we calculate the variances. From the formal solution of $x(t)$ and
$v(t)$, the explicit expressions for the variances are obtained which are
given below;
\begin{eqnarray}
\label{eq25}
\sigma_{xx}^2 (t) & = & \langle\langle [ x (t) - 
\langle\langle x (t) \rangle \rangle ]^2 \rangle\rangle
\nonumber \\
& = & 2 \int_0^t dt_1 \; h (t_1) \int_0^{t_1} dt_2 \; h (t_2) \; 
\langle f(t_1) f(t_2) \rangle \nonumber \\
& & + 2 \left ( \frac{\kappa_0}{g_0} \tau_c \omega_0^2 \right )^2
\int_0^t dt_1 \; h (t_1) \int_0^{t_1} dt_2 \; h (t_2) \; 
\langle \epsilon (t_1) \epsilon (t_2) \rangle_e \nonumber \\
& & + 2 \left ( \frac{\kappa_0}{g_0} \tau_c \right )^2
\int_0^t dt_1 \; h_2 (t_1) \int_0^{t_1} dt_2 \; h_2 (t_2) \; 
\langle \epsilon (t_1) \epsilon (t_2) \rangle_e \nonumber \\
& & + 2 \left ( \frac{\kappa_0}{g_0} \tau_c \right )^2 \omega_0^2 \;
\int_0^t dt_1 \; h (t_1) \int_0^{t_1} dt_2 \; h_2 (t_2) \; 
\langle \epsilon (t_1) \epsilon (t_2) \rangle_e \; \; ,
\end{eqnarray}

\begin{eqnarray}
\label{eq26}
\sigma_{vv}^2 (t) & = & \langle\langle [ v (t) - 
\langle\langle v (t) \rangle\rangle ]^2 \rangle\rangle
\nonumber \\
& = & 2 \int_0^t dt_1 \; h_1 (t_1) \int_0^{t_1} dt_2 \; h_1 (t_2) \; 
\langle f(t_1) f(t_2) \rangle \nonumber \\
& & + 2 \left ( \frac{\kappa_0}{g_0} \tau_c \omega_0^2 \right )^2
\int_0^t dt_1 \; h_1 (t_1) \int_0^{t_1} dt_2 \; h_1 (t_2) \; 
\langle \epsilon (t_1) \epsilon (t_2) \rangle_e \nonumber \\
& & + 2 \left ( \frac{\kappa_0}{g_0} \tau_c \right )^2
\int_0^t dt_1 \; h_3 (t_1) \int_0^{t_1} dt_2 \; h_3 (t_2) \; 
\langle \epsilon (t_1) \epsilon (t_2) \rangle_e \nonumber \\
& & + 2 \left ( \frac{\kappa_0}{g_0} \tau_c \right )^2 \omega_0^2 \;
\int_0^t dt_1 \; h_1 (t_1) \int_0^{t_1} dt_2 \; h_3 (t_2) \; 
\langle \epsilon (t_1) \epsilon (t_2) \rangle_e \; \; ,
\end{eqnarray}

\noindent
and
\begin{eqnarray}
\label{eq27}
\sigma_{xv}^2 (t)  & = & 
\langle\langle [ x (t) - \langle\langle x (t) \rangle\rangle ]
[ v (t) - \langle\langle v (t) \rangle\rangle ] \rangle\rangle \nonumber \\
& = & \frac{1}{2} \dot{\sigma}_{xx}^2 (t) 
\end{eqnarray}

\noindent
where we have assumed that the noises $f(t)$ and $\epsilon (t)$ are symmetric 
with respect to the time argument and have made use the fact that $f(t)$
and $\epsilon (t)$ are uncorrelated.

Due to the Gaussian property of the noises $f(t)$ and $\epsilon (t)$ and the 
linearity of the Langevin equation (\ref{eq15}), we see that the joint
probability density $p(x,v,t)$ of the oscillator must be Gaussian. The joint
characteristic function associated with the density is
\begin{equation}
\label{eq28}
\tilde{p} (\mu,\rho, t)  =  \exp \left \{ i 
\langle\langle x(t) \rangle\rangle \mu +
i \langle \langle v(t) \rangle \rangle \rho -
\frac{1}{2} \left [ \sigma_{xx}^2 (t) \mu^2 +
2 \sigma_{xv}^2 (t) \rho \mu + \sigma_{vv}^2 (t) \rho^2 \right ] \right \}
\; \; .
\end{equation}

\noindent
Using the method of characteristic function \cite{Adelman,Mazo}
and the above expression 
(\ref{eq28}) we find the general Fokker-Planck equation associated with the 
probability density function $p(x,v,t)$ for the process (\ref{eq15});
\begin{equation}
\label{eq29}
\frac{\partial p}{\partial t} = -v \frac{\partial p}{\partial x} +
\bar{\omega}_0^2 (t) x \frac{\partial p}{\partial v} + \bar{\gamma} (t)
\frac{\partial}{\partial v} (vp) + \phi (t) \frac{\partial^2 p}{\partial v^2}
+ \psi (t) \frac{\partial^2 p}{\partial v \partial x} \; \; ,
\end{equation}

\noindent
where
\begin{eqnarray*}
\bar{\gamma} (t) & = & -\frac{d}{dt} \ln \Upsilon (t) \; \; , \; \;
\bar{\omega}_0^2 (t) = \frac{-h(t) \; h_1 (t) + h_1^2 (t)}{\Upsilon (t)}
\; \; {\rm and} \\
\Upsilon (t) & = & \frac{h_1 (t)}{\omega_0^2} \left [ 1 - \omega_0^2
\int_0^\tau d\tau \; h (\tau ) \right ] + h^2 (t) \; \; .
\end{eqnarray*}

\noindent
The functions $\phi (t)$ and $\psi (t)$ are defined by
\begin{equation}
\label{eq30}
\phi (t) = \bar{\omega}_0^2 (t) \sigma_{xv}^2 + \bar{\gamma} 
\sigma_{vv}^2 + \frac{1}{2} \dot{\sigma}_{xv}^2 \; \; {\rm and} \;  \;
\psi (t) = \dot{\sigma}_{xv}^2 + \bar{\gamma} (t) \sigma_{xv}^2 +
\bar{\omega}_0^2 \sigma_{xx}^2 - \sigma_{vv}^2
\end{equation}

\noindent
where the covariances are to be calculated for a particular given noise 
process.

For the internal noise processes it had been shown earlier
that for several models the various time dependent parameters 
$\bar{\omega}_0^2 (t)$, $\bar{\gamma} (t)$, etc. do exist asymptotically as
$t \rightarrow \infty$. The above consideration shows that 
$h(t)$, $h_1(t)$, etc. do not depend on the nature of the noise but
depend only on the relaxation $\bar{\gamma} (t)$. 

We now discuss the asymptotic properties of $\phi (t)$ and $\psi (t)$,
which in turn are dependent on the variances $\sigma_{xx}^2 (t)$ and
$\sigma_{vv}^2 (t)$, as $t \rightarrow \infty$ since they play a significant
role in our further analysis that follows.

\noindent
From Eqs. (\ref{eq25}) and (\ref{eq26}), we may write
\begin{eqnarray*}
\sigma_{xx}^2 (t) = \sigma_{xx}^{2(i)} (t) + \sigma_{xx}^{2(e)} (t) 
\; \; {\rm and} \; \;
\sigma_{vv}^2 (t) = \sigma_{vv}^{2(i)} (t) + \sigma_{vv}^{2(e)} (t) \; \; .
\end{eqnarray*}

\noindent
where `$i$' denotes the part corresponding to internal noise $f(t)$ and
`$e$' corresponds to the external noise $\epsilon (t)$. Since the average
velocity of the oscillator is zero as $t \rightarrow \infty$ we see from
Eq.(\ref{eq22}) that $h (t)$ and $h_1 (t)$ must be zero as 
$t \rightarrow \infty$. Also from Eq.(\ref{eq17}) we observe that the function
$\chi_x (t)$ must decay to zero for long times. Hence, from Eq.(\ref{eq18})
we see that the stationary value of the integral of $h(t)$ is $1/\omega_0^2$,
i.e., 
\begin{equation}
\label{eq30a}
\int_0^\infty h (t) \; dt = \frac{1}{\omega_0^2} \; \; .
\end{equation}

\noindent
Now, $\sigma_{xx}^{2(i)} (t)$ and $\sigma_{vv}^{2(i)} (t)$ of Eqs.(\ref{eq25})
and (\ref{eq26}) can be written in the form
\begin{eqnarray}
\sigma_{xx}^{2(i)} (t) & = & 2 \int_0^t dt_1 \; h (t_1) \int_0^{t_1} dt_2 \;
h (t_2) \; \langle f (t_1) f (t_2) \rangle \nonumber \\
& = & k_BT \left [ 2 \int_0^t d\tau \; h (\tau ) - h^2 (t) -
\omega_0^2 \left \{ \int_0^t d\tau \; h (\tau ) \right \}^2 \right ]
\label{eq30b}
\end{eqnarray}

\noindent
and
\begin{eqnarray}
\sigma_{vv}^{2(i)} (t) & = & 2 \int_0^t dt_1 \; h_1 (t_1) \int_0^{t_1} dt_2 \;
h_1 (t_2) \; \langle f (t_1) f (t_2) \rangle \nonumber \\
& = & k_BT \left [ 1 - h_1^2 (t) - \omega_0^2 h^2 (t) \right ] \; \; .
\label{eq30c}
\end{eqnarray}

\noindent
From the above two expressions [ Eqs.(\ref{eq30b}) and (\ref{eq30c}) ]
we see that 
\begin{equation}
\label{eq30d}
\sigma_{xx}^{2(i)} (\infty) = \frac{k_BT}{\omega_0^2} \; \; \;
{\rm and} \; \; \; \sigma_{vv}^{2(i)} (\infty) = k_BT \; \; .
\end{equation}

\noindent
It is important to note that these stationary values are not related to the 
intensity and correlation time of the internal noise.

We next consider the parts,  $\sigma_{xx}^{2(e)} (t)$ and 
$\sigma_{vv}^{2(e)} (t)$, due to the presence of the external noise. The
Laplace transform of Eq.(\ref{eq16}) yields the expression
\begin{equation}
\label{eq30e}
\tilde{x} (s) - \langle\langle \tilde{x} (s) \rangle \rangle = 
\tilde{h} (s) \tilde{f} (s)
- \frac{\kappa_0}{g_0} \tau_c \omega_0^2 \tilde{h} (s) \tilde{\epsilon} (s)
- \frac{\kappa_0}{g_0} \tau_c s^2 \tilde{h} (s) \tilde{\epsilon} (s)
\end{equation}

\noindent
where
\begin{eqnarray}
\langle\langle \tilde{x} (s) \rangle \rangle & = & \left \{ \frac{1}{s} - 
\frac{\omega_0^2}{s [ s^2 + s \tilde{\gamma} (s) + \omega_0^2] } \right \}
x (0) + \frac{1}{s^2 + s \tilde{\gamma} (s) + \omega_0^2 } \; v (0)
\nonumber \\
& = & \left \{ \frac{1}{s} - \omega_0^2 \frac{ \tilde{h} (s)}{s} \right \}
x (0) + \tilde{h} (s) v (0) \; \; .
\label{eq30f}
\end{eqnarray}

\noindent
From the above equation (\ref{eq30e}) we can calculate the variance
$\sigma_{xx}^2$ in the Laplace-transformed  space which can be
identified as the Laplace 
transform of Eq.(\ref{eq25}). Thus, for the part $\sigma_{xx}^{2(e)} (t)$
we observe that,
$\tilde{\sigma}_{xx}^{2 (e)} (s)$ contains terms like
$\left ( \frac{\kappa_0}{g_0} \tau_c \omega_0^2 \tilde{h} (s) \right )^2 
\langle \tilde{\epsilon}^2 (s) \rangle_e$. Since, we have assumed the
stationarity of the noise $\epsilon (t)$, we conclude that if 
$\tilde{C} (0)$ exists [ where 
$C (t-t') = \langle \epsilon (t) \epsilon (t') \rangle_e $ ],  then the stationary
value of $\sigma_{xx}^{2(e)} (t)$ exists and becomes a constant that
depends on the correlation time and the strength of the noise. Similar
argument is also valid for $\sigma_{vv}^{2(e)} (t)$.

Summarizing the above discussions we note that,

\noindent
(i) the internal noise-driven parts of $\sigma_{xx}^2 (t)$ and 
$\sigma_{vv}^2 (t)$, i.e., $\sigma_{xx}^{2(i)} $ and $\sigma_{vv}^{2(i)}$
approach the fixed values which are independent of the noise correlation and 
the intensity as $t \rightarrow \infty$, 

\noindent
(ii) the external noise driven parts of variances also approach the
constant values at the stationary ($t \rightarrow \infty$) limit which are
dependent on the strength and the correlation time of the noise.

Hence we conclude, following the Ref.(33) and our preceding discussions
that even in presence of an
external noise the above terms do exist asymptotically and we write the
steady state Fokker-Planck equation for the asymptotic values of the
parameters as,
\begin{equation}
\label{eq31}
-v \frac{\partial p}{\partial x} +
\bar{\omega}_0^2 x \frac{\partial p}{\partial v} + \bar{\gamma}
\frac{\partial}{\partial v} (vp) + \phi (\infty) \frac{\partial^2 p}{\partial v^2}
+ \psi (\infty) \frac{\partial^2 p}{\partial v \partial x} = 0 \; \; ,
\end{equation}

\noindent
where, $\bar{\omega}_0^2$, $\bar{\gamma}$, $\phi (\infty )$, $\psi (\infty )$,
etc. are to be calculated from the general definition (\ref{eq30}) for the
steady state. As an explicit example we consider the case
of a $\delta$-correlated external noise and Ornstein-Uhlenbeck internal
noise for which we provide the expressions for variances
$\sigma_{xx}^2 (t)$, $\sigma_{vv}^2 (t)$ and $\sigma_{xv}^2 (t)$ and the 
relaxation function $h(t)$ given in the Appendix-A.

The general steady state solution of the above equation (\ref{eq31}) is
\begin{equation}
\label{eq32}
p_{st} (x,v) = \frac{1}{Z} \exp \left [ - \left \{ \frac{v^2}{2D_0} +
\frac{\bar{\omega}_0^2 x}{ 2 ( D_0 + \psi (\infty) \; )} \right \} \right ]
\end{equation}

\noindent
where
\begin{equation}
\label{eq33}
D_0 = \frac{ \phi (\infty) }{ \bar{\gamma} }
\end{equation}

\noindent
and $Z$ is the normalization constant. The solution (\ref{eq32}) can be
verified by direct substitution. The distribution (\ref{eq32}) is not an 
equilibrium distribution. This stationary distribution for the 
nonequilibrium open system 
plays the role of an equilibrium distribution of the closed system which
may, however, be recovered in the absence of external noise term. 

\section{ Kramers' escape rate
\label{sd4}
}

We now turn to the problem of decay of a metastable state. In Kramers 
approach \cite{Kramers}, the particle coordinate $x$ corresponds to the reaction 
coordinate and its values at the minima of the potential $V(x)$ denotes the 
reactant and product states. 

Linearizing the motion around barrier top at $x=x_b$ the Langevin equation 
(\ref{eq4}) can be written down as
\begin{eqnarray}
\label{eq34}
\dot{y} & = & v \; \; ,\nonumber \\
\dot{v} & = & \omega_b^2 \; y - \int_0^t dt' \; \gamma (t-t') \; v(t') 
+ f (t) + \pi (t) \; \; ,
\end{eqnarray}

\noindent
where, $y=x-x_b$ and the barrier frequency $\omega_b^2$ is defined by
\begin{equation}
\label{eq35}
V (y) = V_b - \frac{1}{2} \omega_b^2 y^2 \; \; ; \; \; \omega_b^2 > 0
\; \; .
\end{equation}

\noindent
Correspondingly the motion of the particle is governed by the 
Fokker-Planck equation (\ref{eq29})
\begin{equation}
\label{eq36}
\frac{\partial p}{\partial t} = -v \frac{\partial p}{\partial y} -
\bar{\omega}_b^2 (t) y \frac{\partial p}{\partial v} + \bar{\gamma}_b (t)
\frac{\partial}{\partial v} (vp) + \phi_b (t) \frac{\partial^2 p}{\partial v^2}
+ \psi_b (t) \frac{\partial^2 p}{\partial v \partial y} \; \; ,
\end{equation}

\noindent
where, the suffix `$b$' indicates that all the coefficients are to be 
calculated using the general definition (\ref{eq30}) for the barrier top 
region.

It is apparent from Eqs.(\ref{eq31}) and (\ref{eq36}) that since the dynamics
is non-Markovian and the system is thermodynamically open one has to deal
with the renormalized frequencies $\bar{\omega}_0$ and $\bar{\omega}_b$ near
the bottom or top of the well, respectively. We
make the ansatz that the nonequilibrium, steady state probability 
$p_b $, generating a nonvanishing diffusion current $j$, across the 
barrier is given by
\begin{equation}
\label{eq37}
p_b (x,v) = \exp \left [ - \left \{ \frac{v^2}{2D_b} +
\frac{ \tilde{V} (x) }{ D_b + \psi_b (\infty) } \right \} \right ]
\; \xi (x,v)
\end{equation}

\noindent
where
\begin{equation}
\label{eq38}
D_b = \frac{ \phi_b (\infty ) }{\bar{\gamma}_b} \; \; .
\end{equation}

\noindent
$\tilde{V} (x)$ is the renormalized linear potential as
\begin{eqnarray}
\label{eq39}
\tilde{V} (x) & = & V (x_0) + \frac{1}{2}\bar{\omega}_0^2 (x-x_0)^2 \; \; ,
\; \; {\rm near \; the \; bottom} \nonumber \\
\tilde{V} (x) & = & V (x_b) - \frac{1}{2}\bar{\omega}_b^2 (x-x_b)^2 \; \; ,
\; \; {\rm near \; the \; top} 
\end{eqnarray}

\noindent
with $\bar{\omega}_0^2$, $\bar{\omega}_b^2 > 0$. The unknown function 
$\xi (x,v)$ obeys the natural boundary condition that for 
$x \rightarrow \infty$, $\xi (x,v)$ vanishes.

The ansatz of the form (\ref{eq37}) denoting the steady state distribution
is motivated by the local analysis near the bottom and the top of the
barrier in the Kramers' sense \cite{Kramers}. For a stationary nonequilibrium
system, on the other hand, the relative population of the two regions, in 
general, depends on the global properties of the potential leading to an
additional factor in the rate expression. Although because of the Kramers'
type ansatz \cite{Kramers}which is valid for the local analysis, such a 
consideration is outside the scope of the present treatment, we point out 
a distinctive feature in the ansatz (\ref{eq37}) compared to Kramers' ansatz.
While in the latter case one considers a complete factorization of the 
equilibrium
part (Boltzmann) and the dynamical part, the ansatz (\ref{eq37}) incorporates
the additional dynamical contribution through dissipation and strength of
the noise into the exponential part. This modification of Kramers' ansatz
(by dynamics) is due to nonequilibrium nature of the system.
Thus unlike Kramers', the exponential
factors in (\ref{eq37}) and in the stationary distribution (\ref{eq32})
which serves as a boundary condition are markedly different. 
Before carrying out global analysis in the present section our aim here is to
understand the modification of the rate due to modulation of the bath
driven by an external noise, within the perview of Kramers' type ansatz. The
internal consistency of the treatment, however, can be checked by recovering
the Kramers' result when the external noise is switched off.

From equation (\ref{eq36}), using (\ref{eq37}) we obtain the equation for 
$\xi (y,v)$ in the steady state in the neighborhood of $x_b$, the equation
\begin{equation}
\label{eq40}
-\left ( 1+ \frac{\psi_b (\infty) }{D_b} \right ) v 
\frac{\partial \xi}{\partial y}
- \left [ \frac{D_b}{D_b+\psi_b (\infty)} 
\bar{\omega}_b^2 y + \bar{\gamma}_b v \right ] 
\frac{\partial \xi}{\partial v} + 
\phi_b (\infty ) \frac{\partial^2 \xi}{\partial v^2} +
\psi_b (\infty ) \frac{\partial^2 \xi}{\partial v \partial y} = 0 \; \; .
\end{equation}

\noindent
After making use of the appropriate transformations and boundary conditions 
for reduced distribution functions\cite{Banik} we obtain the barrier crossing 
rate $k$ given by
\begin{equation}
\label{eq54}
k = \frac{ \bar{\omega}_0 }{2\pi } \; 
\frac{ D_b }{ \{ D_0 + \psi (\infty) \}^{1/2} } \;
\left ( \frac{\Lambda}{1+ \Lambda D_b} \right )^{1/2} \; 
\exp \left [ \frac{ - E_0 }{ D_b + \psi_b (\infty) } \right ]
\end{equation}

\noindent
where 
\begin{eqnarray*}
\Lambda = \frac{\lambda}{\phi_b (\infty) + a \psi_b (\infty)}
\end{eqnarray*}

\noindent
with
\begin{eqnarray*}
a = \frac{D_b}{2 ( D_b + \psi_b (\infty) )} \left \{ -\bar{\gamma}_b
-\sqrt{ \bar{\gamma}_b^2 + 4 \bar{\omega}_b^2 } \right \} \; \; {\rm and}
\; \; \lambda = -\bar{\gamma}_b - a \left ( 1 + 
\frac{\psi_b(\infty)}{D_b} \right ) \; \; .
\end{eqnarray*}

\noindent
Here $E_0$ is the activation energy, $E_0 = V(x_b) - V(x_0)$. Since the
temperature due to internal thermal noise, the strength of the external 
noise and the damping constant are buried in the parameters $D_0$, $D_b$,
$\psi_0$, $\psi_0$ and $\Lambda$ the generalized expression look somewhat 
cumbersome. We point out that the subscripts `$0$' and `$b$' in $D$ and
$\psi$ refer to the well or barrier top region, respectively. Eq.(\ref{eq54})
is one of the key results of this paper. We note here that 
$(D_b + \psi_b (\infty) )/k_B$ in the exponential factor defines a new 
{\it effective} temperature characteristic of the steady state of the
nonequilibrium open system and an {\it effective} transmission factor 
is contained in the
prefactor controlling the barrier crossing dynamics. As expected both are the 
functions of the external noise strength $D$ and coupling of noise to the
bath modes. The dependence of the rate on the parameters can be exposed 
explicitly once we consider the typical cases.

\section{ Numerical analysis of the model and the barrier crossing rate 
\label{sd5}
}

\subsection{ The model : bath modulation vs. direct driving}

We first explore the distinctive aspects of the external 
noise-driven-reservoir-modulated dynamics of the system 
( in contact with the external noise driven bath ) in contrast to
direct driving of the system by the external noise. This will help us to 
elucidate the special role of the reservoir response function in controlling
the rate. To this end we first return to our basic equation of the model,
Eq.(\ref{eq4}) where $\varphi (t-t')$ as given by Eq.(\ref{eq9}) characterizes
the response function. We solve this equation numerically using second order
stochastic algorithm of Fox \cite{Fox} for the full potential
$V(x) = \frac{1}{4} x^4 - \frac{1}{2} x^2$ and plot the results of 
computation of the inverse of the mean first passage time as a function
of the damping constant $g_0^2$ ( see Eq.(\ref{eq12}) ) in Fig.(1). The 
density of 
modes of the bath has been assumed to be of the standard Lorentzian form,
${\cal D} (\omega) = \frac{2}{\pi} \frac{1}{\tau_c} \left ( 
\frac{1}{\omega^2 +\tau_c^{-2}} \right )$
with vanishing internal correlation time $\tau_c$. Typically for the other
scaled parameters we have used $k_BT$ = 0.1, the strength of the
$\delta$-correlated external noise, $D$ = 0.1. The mean first passage times
have been calculated by averaging over 10000 trajectories. We first show the 
continuous curve in Fig.(1) which represents the case of direct driving
by the external $\delta$-correlated noise. What is immediately apparent is the
loss of turnover of the rate as one allows the variation of the dissipation 
constant $g_0^2$
from low to high friction regime. This is in sharp contrast to what
is shown by the dotted curve in Fig.(1) which depicts the situation of bath
modulation with the response function $\varphi (t-t')$ determined by 
Eq.(\ref{eq13}). Both the curves ( continuous and dotted )
representing the open system are compared to that for
the standard Kramers' turnover, i.e., when the dynamics is calculated
in absence of the external noise ( dashed curve ). 
As expected the rate in the driven system
( be it directly or through bath modulation ) is always higher that in the
undriven one. Fig.(1) also shows that although in the low damping
region the direct driving causes a much higher rate, it is, however, the
bath modulation which becomes more effective in inducing activated barrier
crossing in the high friction regime. It is thus interesting to note that
the barrier crossing dynamics of the system in contact with an external 
noise modulated bath captures the essential turnover features of the Kramers'
dynamics of the closed system. We therefore realize that although open,
the nature of the response function of the reservoir as well as the 
thermodynamic consistency condition (\ref{neq6}), make the open system feel
like a closed system.

In Fig.(2) we show the variation of inverse of the mean first passage time
as a function of the strength of the external noise $D$, keeping all other 
parameter same as before but for $g_0^2$ = 2.0. It is apparent that the
barrier crossing is more facilitated by modulating the bath than driving the
system directly for higher values of external noise strength.

\subsection{The rate : Theoretical results vs. numerical simulation}

So far we have considered the full potential 
$V(x) = \frac{1}{4} x^4 - \frac{1}{2} x^2$ and vanishingly small correlation
times for external and internal noises for numerical computation. We now turn 
to our basic theoretical result Eq.(\ref{eq54}) which is a generalization of
Kramers' rate for bath modulated dynamics for intermediate to strong
damping regime. To examine its validity we calculate the rate (\ref{eq54}) as 
a function of the damping constant $g_0^2$ 
( $g_0$ is related to both $\gamma (t)$
and $\varphi (t)$ in Eqs.(\ref{eq12}) and (\ref{eq13}) respectively ) for 
several values of external noise strength $D$.  The scaled barrier height 
$E_0$ and $k_BT$ have been set to 0.25 and 0.1, respectively. The results are 
plotted in Fig.(3) ( continuous curve ) and compared to the rate, inverse of 
the mean first passage time ( dotted curve ) calculated numerically using
full potential for the set of parameter values ( as given in the earlier
subsection ) with Eq.(\ref{eq4}).  In Figs.(4) and (5) we further compare our 
theoretical results ( continuous curve ) with numerical simulation ( dotted
curve ) for the variation of rate as a function of external noise strength
and its correlation time, respectively for several values of damping
constant. It is apparent from Figs.(3-5) that the theoretical and numerical 
results are in good agreement.

\section{ Conclusions 
\label{sd6}
}

Based on a system-reservoir microscopic model where the reservoir is 
modulated by an external, stationary and Gaussian noise with arbitrary
decaying correlation function, we have numerically analyzed the model and
generalized the Kramers' theory
to calculate the steady state rate of escape from a metastable well. The main
conclusions of this study are as follows;

(i) We have shown that since the reservoir is driven by the external noise
and the dissipative properties of the system depend on the reservoir, a
simple connection between the dissipation and the response function of the
medium due to the external noise can be established.
This connection is important for realizing an effective temperature of the 
reservoir characterizing the stationary state of the
thermodynamically open system, as well as an {\it effective transmission 
factor} controlling the rate. Both of these quantities depend on the strength 
and correlation time of the external noise.

(ii) Many of the earlier treatments of the rate concern direct
(phenomenological) driving of the system and did not emphasize
the question of thermodynamic consistency in the context of open systems. 
The present theory being microscopic the fluctuation-dissipation-like 
relation (\ref{neq6}) remains an inbuilt characteristic of the model itself 
as an essential thermodynamic consistency condition.

(iii) Based on numerical simulation of the full model potential we show that 
one can recover the turnover features of the Kramers' dynamics when the
external noise modulates the reservoir rather than the system directly. This
recovery, we believe, is an offshoot of the derived thermodynamic consistency
condition (\ref{neq6}).

(iv) Provided the long time limit of the moments for the stochastic processes
pertaining to the external and internal noises characterized by arbitrary
decaying correlation functions exist, the expression for generalized
Kramers' rate of barrier crossing for the open system we derive here is
fairly general. We have shown that it agrees reasonably well with numerical
simulation using full potential for this model.

The creation of a typical nonequilibrium open situation by modulating a bath
with the help of
an external noise is not an uncommon phenomenon in applications and
industrial processing. The external agency generating noise does work on the
bath by stirring, pumping, agitating, etc., to which the system dissipates 
internally. In the present treatment we are concerned with a nonequilibrium
steady state which signifies 
a constant throughput of energy in contrast to thermal equilibrium defined
by an constant temperature. We believe that these considerations are likely
to be important in other related issues in nonequilibrium open systems
and may serve as a basis for studying processes occurring within
irreversibly driven environments \cite{JRC-1,RIG} and for thermal ratchet
problems \cite{Astumian}.

\acknowledgments
We thank E. Pollak for his kind interest in this work 
and for pointing out the Ref.~{(35)} to us.
SKB in indebted to Council of Scientific and Industrial Research (CSIR),
Government of India for financial support.

\begin{appendix}

\section{Calculation of variances}

We consider a particular case as an example where the external noise 
$\epsilon (t)$ is $\delta$-correlated and the internal noise is an
Ornstein-Uhlenbeck process, i.e., 
\begin{eqnarray*}
\langle \epsilon (t) \epsilon (t') \rangle_e = 2 D \delta (t-t')
\end{eqnarray*}

\noindent
and
\begin{eqnarray*}
\langle f (t) f (t') \rangle = \frac{g_0^2 k_BT}{\tau_c}
e^{-|t-t'| / \tau_c} \; \; .
\end{eqnarray*}

\noindent
Consequently, from the fluctuation-dissipation relation we derive the 
dissipative kernel as,
\begin{eqnarray*}
\gamma (t-t') = \frac{g_0^2 }{\tau_c} e^{-|t-t'| / \tau_c} \; \; .
\end{eqnarray*}

\noindent
It should be noted that for $\tau_c \rightarrow 0$, the above noise
process become $\delta$-correlated.

The Laplace transform of $\gamma (t)$ as given above can be written as
\begin{eqnarray*}
\tilde{\gamma} (s) = \frac{ g_0^2 }{s \tau_c +1 } \; \; ,
\end{eqnarray*}

\noindent
and subsequently, we have for $\tau_c \neq 0$, 
\begin{eqnarray*}
\tilde{h} (s) = \frac{ s+a }{
s^3 + a s^2 + b s + c_0}
\end{eqnarray*}

\noindent
where
\begin{eqnarray*}
a = \frac{1}{\tau_c} \; \; , \; \; b = \omega_0^2 + \frac{g_0^2}{\tau_c}
\; \; {\rm and} \; \;  c_0 = \frac{\omega_0^2 }{\tau_c} \; \; .
\end{eqnarray*}

We find that the inverse Laplace transform of $\tilde{h}(s)$ reads
\begin{equation}
\label{eqa1}
h (t) = c_1 e^{-\Delta_1 t} + c_2 e^{-\Delta_2 t} \; \sin(\beta t + \alpha)
\end{equation}

\noindent
where the coefficients $c_1$, $c_2$, $\Delta_1$, $\Delta_2$, $\beta$ and $\alpha$ are 
given by
\begin{mathletters}
\begin{eqnarray}
\Delta_1 & = & - {\cal A} - {\cal B} + \frac{a}{3} \; \; ,\\
\Delta_2 & = & \frac{1}{2}({\cal A} + {\cal B}) + \frac{a}{3} \; \; ,\\
\beta & = & \frac{\sqrt{3} }{2} ( {\cal A} - {\cal B} ) \; \; ,\\
c_1 & = & \frac{1}{ 2\Delta_2 - \Delta_1 - d } \; \; ,\\
d & = & \frac{ a (2\Delta_2 - \Delta_1) - \Delta_2^2 -\beta^2 }{ a - \Delta_1 } \; \; ,\\
{\cal A} & = & \left ( -\frac{a^3}{27} + \frac{ab}{6} - \frac{c_0}{2} + \sqrt{{\cal Q}}
\right )^{1/3} \; \; ,\\
{\cal B} & = & \left ( -\frac{a^3}{27} + \frac{ab}{6} - \frac{c_0}{2} - \sqrt{{\cal Q}}
\right )^{1/3} \; \; ,\\
c_2 & = & -\frac{c_1}{\beta} [ (d-\Delta_2)^2 + \beta^2 ]^{1/2} \; \; ,\\
\alpha & = & \tan^{-1} \left ( \frac{\beta}{d-\Delta_2} \right ) \; \;
{\rm and}\\
{\cal Q} & = & -\frac{a^2b^2}{108} + \frac{b^3}{27} + \frac{a^3 c_0}{27}
- \frac{abc_0}{6} + \frac{c_0^2}{4} \; \; .
\end{eqnarray}
\end{mathletters}

\noindent
Here we note that for a physically allowed solution $\Delta_1$, $\Delta_2$
must be positive. Since by Eq.(\ref{eq19}) $h(t)$ depends on the memory
kernel $\gamma (t)$ which is of decaying type and all the moments, in
general, reach asymptotic constancy as shown in Sec.~{III}, these quantities
are positive (which depends on the correlation time $\tau_c$, the strength
of the noise and other potential parameters) which may be checked (after
some algebra) by considering the limiting cases such as $\tau_c \rightarrow 0$
and $\tau_c \rightarrow$ large.

Substituting Eq.(\ref{eqa1}) into the expressions for variances [external
noise is $\delta$-correlated], namely into (\ref{eq25}) and (\ref{eq26})
we have after some lengthy algebra
\begin{eqnarray*}
\sigma_{xx}^2 (t) = \sigma_{xx}^{2(i)} (t) + \sigma_{xx}^{2(e)} (t)
\end{eqnarray*}

\noindent
where
\begin{eqnarray}
\sigma_{xx}^{2(i)} (t) & = & k_BT \left ( c_2 R + \frac{c_1}{\Delta_1} \right )
\left [ 2 - \omega_0^2 \left ( c_2 R + \frac{c_1}{\Delta_1} \right ) \right ]
\nonumber \\
& & + k_BT \left \{ -\frac{c_1}{\Delta_1} e^{-\Delta_1 t} \left [ 2 - 
2 \omega_0^2 c_2 R - \frac{2 \omega_0^2 c_1}{\Delta_1} + e^{-\Delta_1t}
\left ( \Delta_1 c_1 + \frac{\omega_0^2 c_1}{\Delta_1} \right ) \right ] \right.
\nonumber \\
& & - \frac{2 c_2 e^{-\Delta_2 t} }{\Delta_2^2 + \beta^2} \left [ 1 -
\omega_0^2 c_2 R + \frac{\omega_0^2 c_1}{\Delta_1} ( e^{-\Delta_1 t} - 1 )
\right ] \left [ \Delta_2 \sin (\beta t + \alpha) + \beta \cos (\beta t + \alpha)
\right ] \nonumber \\
& & - 2 c_1 c_2 e^{ - (\Delta_1+\Delta_2) t} \sin (\beta t + \alpha) \nonumber \\
& & - \frac{\Delta_2 \beta \omega_0^2 c_2^2 e^{-2\Delta_2t} }{ (\Delta_2^2+\beta^2)^2 }
\sin 2 (\beta t + \alpha)
- \frac{\beta^2 \omega_0^2 c_2^2 e^{-2\Delta_2t} }{ (\Delta_2^2+\beta^2)^2 }
\nonumber \\
& & \left. + \left [ 
\frac{ \omega_0^2 (2\beta^2 - \Delta_2^2) }{ (\Delta_2^2+\beta^2)^2 } -1 \right ]
c_2^2 e^{-2\Delta_2t} \sin^2 (\beta t + \alpha ) \right \}
\end{eqnarray}

\noindent
with
\begin{equation}
\label{eqa2}
R = \frac{1}{\Delta_2^2 + \beta^2} ( \Delta_2 \sin \alpha + \beta \cos \alpha)
\end{equation}

\noindent
and
\begin{eqnarray}
\label{eqa3}
\sigma_{xx}^{2(e)} (t) & = & 
2D \left ( \frac{\kappa_0}{g_0} \tau_c \right )^2
[ c_1^2 ( \omega_0^4 + \Delta_1^4 + 2\omega_0^2 \Delta_1^2 ) I_A (t) \nonumber \\
& & + c_2^2 \{ \omega_0^4 + (\Delta_2^2-\beta^2)^2 - 4\beta^2 \Delta_2^2 + 
2 \omega_0^2 ( \Delta_2^2 - \beta^2) \} I_B (t) \nonumber \\
& & + 2 c_1 c_2 \{ \omega_0^4 + \Delta_1^2 (\Delta_2^2-\beta^2)  
+ \omega_0^2 ( \Delta_1^2 + \Delta_2^2 - \beta^2) \} I_C (t) \nonumber \\
& & - 2 c_2^2 \beta \Delta_2 ( \Delta_2^2 -\beta^2 + \omega_0^2 ) I_D (t)
+ 4 c_2^2 \beta^2 \Delta_2^2 I_E (t) 
- 4 c_1 c_2 \beta \Delta_2 ( \Delta_1^2 + \omega_0^2 ) I_F (t) ] \; \; .
\end{eqnarray}

\noindent
Here the $I$'s are defined by
\begin{mathletters}
\begin{eqnarray}
\label{eqa4}
I_A (t) & = & \int_0^t e^{-2\Delta_1t} \; dt \; \; , \\
I_B (t) & = & \int_0^t e^{-2\Delta_2t} \sin^2 (\beta t +\alpha) \; dt \; \; , \\
I_C (t) & = & \int_0^t e^{-(\Delta_1+\Delta_2)t} \sin (\beta t +\alpha) \; dt \; \; , \\
I_D (t) & = & \int_0^t e^{-2\Delta_2t} \sin 2(\beta t +\alpha) \; dt \; \; , \\
I_E (t) & = & \int_0^t e^{-2\Delta_2t} \; dt \; \; {\rm and} \\
\label{eqa5}
I_F (t) & = & \int_0^t e^{-(\Delta_1+\Delta_2)t} \cos (\beta t +\alpha) \; dt \; \; .
\end{eqnarray}
\end{mathletters}

\noindent
Similarly
\begin{eqnarray*}
\sigma_{vv}^2 (t) = \sigma_{vv}^{2(i)} (t) + \sigma_{vv}^{2(e)} (t)
\end{eqnarray*}

\noindent
where
\begin{eqnarray}
\sigma_{vv}^{2(i)} (t) & = & k_BT - [ ( \Delta_1^2 + \omega_0^2 ) c_1^2 e^{-2\Delta_1t}
+ \beta^2 c_2^2 e^{-2\Delta_2t} \nonumber \\
& & - \beta \Delta_2 c_2^2 e^{-2\Delta_2t} \sin 2 (\beta t + \alpha ) +
( \Delta_2^2 + \omega_0^2 - \beta^2 ) c_2^2 e^{-2\Delta_2t}
\sin^2 (\beta t + \alpha )\nonumber \\
& & + e^{-(\Delta_1+\Delta_2)t} \{ 2c_1c_2 (\omega_0^2 + 
\Delta_1 \Delta_2 )\sin (\beta t + \alpha )
- 2 \Delta_1 \beta c_1 c_2 \cos (\beta t + \alpha ) \} ]
\end{eqnarray}

\noindent
and
\begin{eqnarray}
\label{eqa6}
\sigma_{vv}^{2(e)} (t) & = & 
2D \left ( \frac{\kappa_0}{g_0} \tau_c \right )^2
[ c_1^2 \Delta_1^2 ( \omega_0^2 + \Delta_1^2 )^2 I_A (t) \nonumber \\
& & + c_2^2 \{ ( \omega_0^2 + \Delta_2^2 - 3\beta^2)^2 \Delta_2^2 
- ( \omega_0^2 + 3 \Delta_2^2 - \beta^2)^2 \beta^2 \} I_B (t) \nonumber \\
& & + 2 c_1 c_2 \Delta_1 \Delta_2 ( \omega_0^2 + \Delta_1^2 )  
( \omega_0^2 - 3 \beta^2 + \Delta_2^2 ) I_C (t) \nonumber \\
& & - c_2^2 \beta \Delta_2 ( 3 \Delta_2^2 -\beta^2 + \omega_0^2 ) 
(\omega_0^2 - 3\beta^2 + \Delta_2^2) I_D (t) \nonumber \\
& & + c_2^2 \beta^2 ( \omega_0^2 -\beta^2 + 3 \Delta_2^2 ) I_E (t) \nonumber \\
& & - 2 c_1 c_2 \beta \Delta_1 ( \Delta_1^2 + \omega_0^2 ) 
( \omega_0^2 + 3 \Delta_2^2 -\beta^2 ) I_F (t) ]
\end{eqnarray}

\noindent
where, $I$'s are defined in Eq.(\ref{eqa4}-\ref{eqa5}). 
The explicit expression for 
$\sigma_{xv}^2 (t)$ can be derived from Eq.(\ref{eq27}).
In the limit $t \rightarrow \infty$ we calculate the stationary values
of the variances.
The variances $\sigma_{xx}^2 (\infty)$, $\sigma_{vv}^2 (\infty)$ and 
$\sigma_{xv}^2 (\infty)$ yield $\phi (\infty)$ and $\psi (\infty)$ and other 
relevant quantities.

\end{appendix}

\begin{figure}
\caption{ Plot of barrier crossing rate, $k$ vs. damping constant, $g_0^2$.
The solid and the dotted line correspond to direct driving of the system 
and bath modulation, respectively ($D$ = 0.1). The dashed line corresponds 
to the thermodynamically closed system, i.e., the system without any external 
driving ($D$ = 0.0). $k_BT$ = 0.1 is common for all the three curves.
(units are arbitrary)
}
\end{figure}

\begin{figure}
\caption{ Plot of barrier crossing rate, $k$ vs. external noise strenth, D 
for a constant $g_0^2$. The curve (a) represents the results for bath
modulation while the curve (b) is the result for direct additive driving.
(units are arbitrary)
}
\end{figure}

\begin{figure}
\caption{ Plot of barrier crossing rate, $k$ vs. damping constant, $g_0^2$
for different external noise strengths $D$. The solid lines correspond to 
theoretical result ( Eq.(52) ) and the dotted curves are due to simulation.
(a) $D$ = 0.15, (b) $D$ = 0.10 and (c) $D$ = 0.05. (units are arbitrary)
}
\end{figure}

\begin{figure}
\caption{ Plot of barrier crossing rate, $k$ vs. external noise strength,
$D$ for different values of $g_0^2$. The solid and the dotted lines are
same as in Fig.(3). (a) $g_0^2$ = 2.0 and (b) $g_0^2$ = 3.0 .
(units are arbitrary)
}
\end{figure}

\begin{figure}
\caption{ Plot of barrier crossing rate, $k$ vs. correlation time of the
external
noise, $\tau_e$ for different values of external noise strengths, $D$. The 
solid and the dotted lines are same as in Fig.(3). (a) $D$ = 0.5, (b) $D$ =
1.0 and (c) $D$ = 1.5 . (units are arbitrary)
}
\end{figure}


\begin{thebibliography}{99}

\bibitem{Kramers} H. A. Kramers, Physica (Amsterdam) {\bf 7}, 284 (1940).

\bibitem{Van} N. G. van Kampen, Prog. Theo. Phys. {\bf 64}, 389 (1978).

\bibitem{Grote-Hynes} R. F. Grote and J. T. Hynes, J. Chem. Phys. {\bf 73}, 2715
(1980).

\bibitem{Hanggi} P. H\"anggi and F. Mojtabai, Phys. Rev. A {\bf 26}, 1168 (1982).

\bibitem{Pollak} E. Pollak, J. Chem. Phys. {\bf 85}, 865 (1986).

\bibitem{JRC-1} J. Ray Chaudhuri, G. Gangopadhyay and D. S. Ray, J. Chem. Phys.
{\bf 109}, 5565 (1998).

\bibitem{Banik} S. K. Banik, J. Ray Chaudhuri and D. S. Ray, J. Chem. Phys.
{\bf 112}, 8330 (2000).

\bibitem{Caldeira} A. O. Caldeira and A. J. Leggett, Phys. Rev. Lett. {\bf 46}, 
211 (1981).

\bibitem{Grabert-1} H. Grabert, P. Schramm and G. L. Ingold, Phys. Rep. 
{\bf 168}, 115 (1988).

\bibitem{Grabert-2} H. Grabert, U. Weiss and P. H\"anggi, 
Phys. Rev. Lett. {\bf 52}, 2193 (1984).

\bibitem{JRC-2} J. Ray Chaudhuri, B. C. Bag and D. S. Ray, J. Chem. Phys.
{\bf 111}, 10852 (1999).

\bibitem{RMP} P. H\"anggi, P. Talkner and M. Borkovec, Rev. Mod. Phys. 
{\bf 62}, 251 (1990).

\bibitem{Melnikov} V. I. Mel'nikov, Phys. Rep. {\bf 209}, 1 (1991).

\bibitem{Talkner} P. Talkner, E. Pollak and A. M. Berezhkovskii, Eds.,
Chem. Phys. {\bf 235}, 1 (1998).

\bibitem{Weiss} U. Weiss, {\it Quantum Dissipative Systems} 
(World Scientific, Singapore, 1993).

\bibitem{Ford} G. W. Ford, M. Kac and P. Mazur, J. Math. Phys. {\bf 6},
504 (1965).

\bibitem{Louisell} W. H. Louisell, {\it Quantum Statistical Properties of
Radiation} (Wiley, New York, 1973).

\bibitem{Zwanzig} R. Zwanzig, J. Stat. Phys. {\bf 9}, 215 (1973).

\bibitem{Lindenberg} K. Lindenberg and V. Seshadri, Physica A {\bf 109}, 483
(1981).

\bibitem{Kubo} R. Kubo, M. Toda and N. Hashitsume, {\it Statistical Physics
II, Nonequilibrium Statistical Mechanics} (Springer, Berlin, 1985).

\bibitem{West} K. Lindenberg and B. J. West, {\it The Nonequilibrium
Statistical Mechanics of Open and Closed Systems} (VCH Publisher, Inc.,
New York, 1990).

\bibitem{Astumian} R. D. Astumian, Science {\bf 276}, 917 (1997) and the
references given therein. 

\bibitem{Rattray} 
K. M. Rattray and A. J. McKane, J. Phys. A {\bf 24},
4375 (1991).

\bibitem{Moss} See, for example,
F. Moss and P. V. E. McClintock, Eds., {\it Noise in nonlinear
dynamical systems}, Vol.~{I-III} (Cambridge University Press, England, 1989).

\bibitem{Jaume} J. Masoliver and J. M. Porr\`a, Phys. Rev. E {\bf 48}, 4309
(1993).

\bibitem{Werner} W. Horsthemke, C. R. Doering, T. S. Ray and M. A. Burschka,
Phys. Rev. A {\bf 45}, 5492 (1992).

\bibitem{SJBE} S. J. B. Einchcomb and A. J. McKane, Phys. Rev. E {\bf 49},
259 (1994).

\bibitem{WH-RL} W. Horsthemke and R. Lefever, {\it Noise-Induced Transitions}
(Springer-Verlag, Berlin, 1984).

\bibitem{Farkas} L. Farkas, Z. Phys. Chem. (Leipzig) {\bf 125}, 236 (1927).

\bibitem{Bravo} J. M. Bravo, R. M. Velasco and J. M. Sancho, J. Math. Phys.
{\bf 30}, 2023 (1989).

\bibitem{Adelman} S. A. Adelman, J. Chem. Phys. {\bf 64}, 124 (1976).

\bibitem{Mazo} R. M. Mazo in {\it Stochastic Processes in Nonequilibrium 
Systems}, edited by L. Garrido, P. Segler and P. J. Shepherd, Lecture Notes
in Physics, Vol.~{84} (Springer-Verlag, Berlin, 1978).

\bibitem{BC-AN} B. Carmeli and A. Nitzan, J. Chem. Phys. {\bf 29}, 1481
(1984).

\bibitem{Fox} R. F. Fox, Phys. Rev. A {\bf 43}, 2649 (1991).

\bibitem{RIG} R. Hernandez, J. Chem. Phys. {\bf 111}, 7701 (1999).

\end{thebibliography}
\end{document}